\documentclass[conference]{IEEEtran}

\usepackage{cite}
\usepackage{amsmath,amssymb,amsfonts}
\usepackage{algorithmic}
\usepackage{graphicx}
\usepackage{textcomp}
\usepackage{xcolor}
\usepackage[a4paper, total={184mm,239mm}]{geometry}
\def\BibTeX{{\rm B\kern-.05em{\sc i\kern-.025em b}\kern-.08em
    T\kern-.1667em\lower.7ex\hbox{E}\kern-.125emX}}

\usepackage{enumitem}
\usepackage{pgfplots}
\usepackage{soul}
\usepackage{svg}

\usepackage[colorlinks=true, linkcolor=blue, citecolor=blue, urlcolor=blue]{hyperref}
\usepackage{booktabs}
\usepackage{multirow}
\usepackage{color}
\usepackage{comment}          
\usepackage{amsthm}
\usepackage{etoolbox}         
\usepackage{subfig}
\usepackage{balance} 
\usepackage{cleveref}
\usepackage{xspace}
\usepackage{pifont}
\usepackage{tabularx}

\usepackage{mathtools}

\usepackage{tabularx,booktabs, threeparttable}
\newcolumntype{Y}{>{\centering\arraybackslash}X}

\usepackage[colorlinks=true, linkcolor=blue, citecolor=blue, urlcolor=blue]{hyperref}

\graphicspath{{./figs/}}

\newif\ifshowmethodrevisions
\showmethodrevisionstrue

\begin{document}
\IEEEoverridecommandlockouts

\title{
Timing-Driven Logic Remapping with Local Physical Context
}

\author{\IEEEauthorblockN{Zijian Jiang, Hongyang Pan, Cunqing Lan, Keren Zhu$^{*}$\thanks{$*$ Corresponding author: krzhu@fudan.edu.cn}}
\IEEEauthorblockA{State Key Laboratory of ASIC and System, Fudan University, Shanghai, China}
}

\maketitle

\begin{abstract}

The timing behavior of a mapped circuit depends on both its logic implementation and the physical environment in which that implementation is realized. Revisiting mapping decisions after placement therefore requires a search procedure that accounts for surrounding timing constraints, fanout loads, and interconnect effects. We study local remapping in this setting and develop a framework that couples discrete mapping search with physical implementation feedback. Timing-critical regions are isolated through bounded windows whose interfaces retain the context of the surrounding circuit. Within each window, a mixed-integer formulation jointly selects logic cuts, signal polarities, and library cells under a delay model informed by estimated locations and interconnect parasitics. A continuous relaxation filters the search space before discrete optimization produces alternative implementations with similar modeled timing and different structural choices. These implementations are reconstructed and assessed through legalization, routing-based parasitic estimation, and timing analysis. Physically validated improvements are incorporated into the design, and the updated context guides subsequent searches. The framework provides a systematic way to revisit local logic implementations while accounting for their interaction with an existing placement.
\end{abstract}


\section{Introduction}
\label{sec:Introduction}

Timing closure remains one of the most iterative stages in physical design. Gate sizing and buffer insertion adjust delay by modifying cell drive strength or inserting repeaters, but do not explore alternative Boolean decompositions. When the original logic decomposition is suboptimal, the existing structure can limit the timing improvements achievable through these operations. Logic restructuring complements this toolkit by replacing combinational regions with functionally equivalent implementations that may offer better timing after physical realization. Applying restructuring to an already placed design provides access to the surrounding physical context, but also requires the new implementation to remain compatible with that context.




Post-placement local restructuring raises two related challenges. First, replacing gates introduces new instances that require placement and legalization. A larger region exposes more optimization opportunities, but may disrupt more of the established placement and alter the loads seen by neighboring gates. Extraction must therefore balance optimization scope against physical disruption. Second, even a bounded region admits a combinatorial space of cut, phase, and cell choices. A mapping cost model, including one informed by physical information, may rank these choices differently from timing analysis after reconstruction, legalization, and parasitic updates. A modeled improvement thus need not survive physical implementation. Evaluating every alternative in the placed design is costly, motivating selective candidate generation followed by physical validation.

Prior efforts connect logic optimization and physical design in several ways. Concurrent approaches interleave logic optimization with placement to account for their interaction~\cite{jiang1997postlayout,gosti2001integrating,lou1999concurrent,stok2001cycletime}. Layout-driven synthesis also considers congestion during decomposition and mapping~\cite{kutzschebauch2001congestion}. Physically aware technology mapping incorporates spatial information into mapping decisions: PigMAP uses primitive-gate placement to guide mapping~\cite{pan2024pigmap}, and PigMap2 extends the use of physical information in the mapping framework~\cite{pan2025pigmap2}. PigMap3 further studies incremental mapping with on-the-fly post-layout critical-path tracking~\cite{pan2026pigmap3}. CPA-Remap constructs local cones using critical paths and cell locations for physically aware remapping~\cite{he_cpa-remap_2025}. These approaches establish the relevance of physical context to mapping optimization. Our focus is a complementary question: how can the timing and load context of a placed design guide joint local mapping decisions while accounting for the physical effects of implementing those decisions?

Our key insight is to guide local restructuring with the timing and load context of the placed design and estimates of candidate interconnect delay, while using physical evaluation to determine candidate acceptance. Based on this insight, we propose Physically Contextualized Restructuring (PCR), which extracts bounded windows around timing-critical endpoints and jointly optimizes cut, phase, and library-cell choices. Reference timing and external pin loads retain the surrounding design context. Predicted gate positions provide wirelength and RC estimates for candidate scoring; these estimates remain fixed within each solve and can be refreshed using the selected topology. A linear programming (LP) relaxation screens windows for modeled improvement and restricts the mapping domain for integer optimization. Sparse and structurally distinct candidates are selected under a common modeled timing cap, reconstructed, and evaluated through legalization and global-routing-based timing analysis. Accepted candidates update the physical design and provide the reference context for subsequent iterations.

Our main contributions are as follows:
\begin{itemize}
\item A local mapping formulation that jointly models cut, phase, and cell choices within bounded windows, incorporating estimated interconnect delays while preserving shared boundary signals and the surrounding timing context (Sections~\ref{sec:extraction}--\ref{sec:lp_mip}).
\item A staged solution procedure that combines relaxation-bound screening and action-domain restriction with integer feasibility and timing optimization (Sections~\ref{sec:lp_mip}--\ref{sec:representatives}).
\item A dual-representative generation strategy that selects sparse and structurally distinct mappings under a common modeled timing cap, enabling separate physical evaluation of alternatives with similar modeled timing (Section~\ref{sec:representatives}).
\item An experimental evaluation showing that the complete PCR flow outperforms PigMap3~\cite{pan2026pigmap3} with 86.7\%/95.2\% reductions in geometric-mean WNS/TNS violation magnitudes under position-aware postprocessing, and genetic and annealing with 48.6--49.2\%/53.3--63.1\% reductions under uniform postprocessing. It outperforms PigMap3 in WNS on seven of nine cases and each resynthesis baseline on six of seven cases, with one tie (Section~\ref{sec:results}).
\end{itemize}

The source code is available at \url{https://github.com/jzj3021/PCR}.

\section{Preliminaries}
\label{sec:related}

\subsection{Logic Synthesis and Technology Mapping}

Logic synthesis optimizes a technology-independent representation, such as an And-Inverter Graph (AIG), before mapping it to library cells. Cut-based mapping~\cite{keutzer1987dagon,chatterjee2006structural} selects compatible implementations covering the required outputs, often from a restricted set of priority cuts~\cite{mishchenko2007priority}. Delay-oriented methods include load-aware ExactMap~\cite{xie_exactmap_2025} and learned delay prediction in AiMap+~\cite{liu_aimap_2024}. PCR uses ABC~\cite{mishchenko2007abc} to obtain legal cut, phase, and cell alternatives and their dependencies, defining the discrete domain optimized in Section~\ref{sec:lp_mip}.

\begin{figure*}
  \centering
  \includegraphics[width=\linewidth]{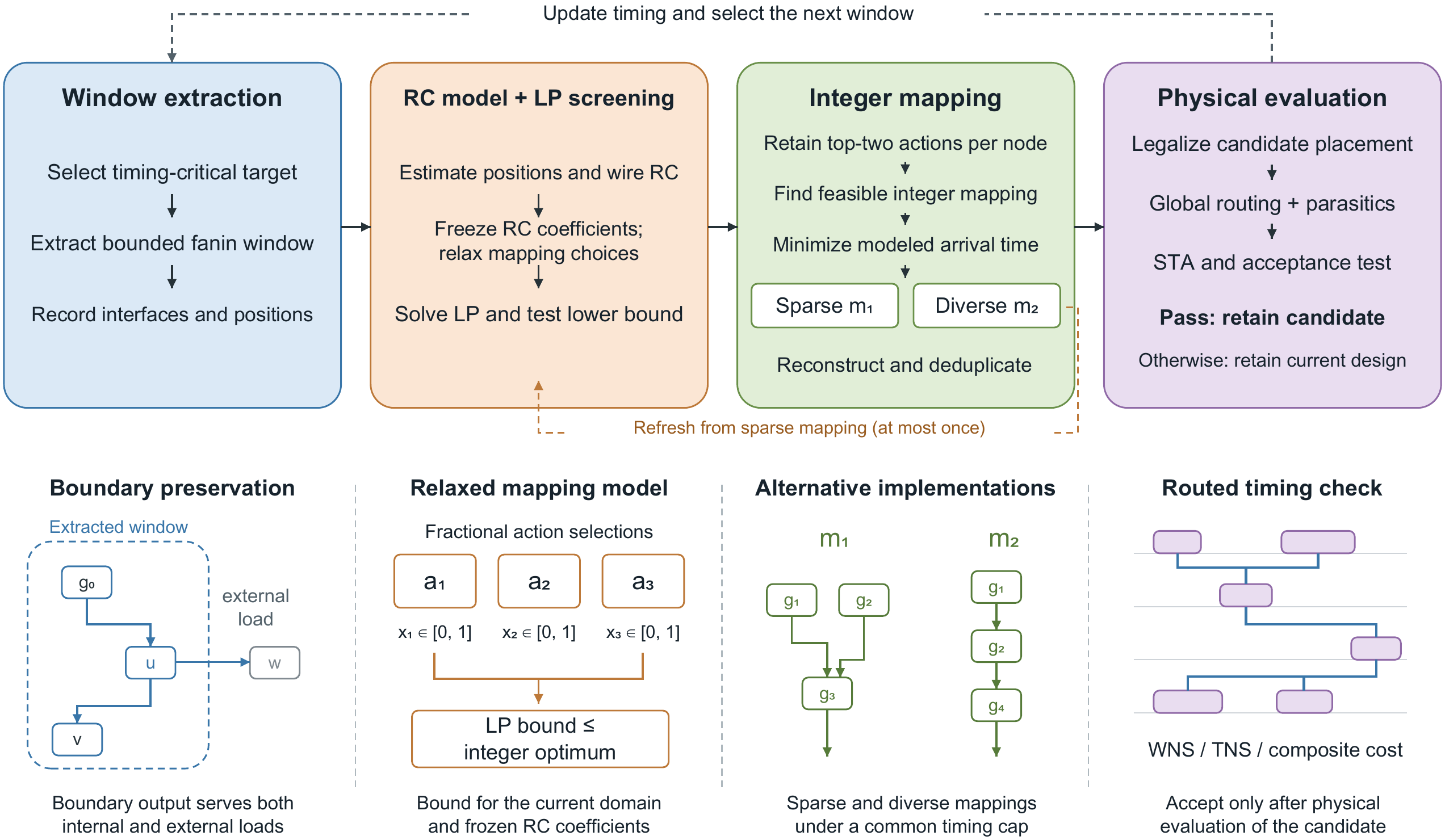}
  \caption{Overview of PCR. Window interfaces provide fixed physical context for candidate position and RC estimation. LP screening and integer optimization use a frozen delay model to generate sparse and diverse mappings. The sparse mapping can refresh the model for one additional solve. Reconstructed candidates undergo legalization, global routing, and timing analysis before acceptance.}
  \label{fig:overview}
\end{figure*}

\subsection{Linear Programming Relaxation}
\label{sec:prelim_lp}

An LP relaxation replaces binary selections $x\in\{0,1\}$ by $x\in[0,1]$ while retaining the linear constraints. Let $\mathcal M$ be the integer feasible set, $\mathcal P\supseteq\mathcal M$ its relaxation, and $\widehat T$ the modeled arrival-time objective. For a feasible minimization problem,

\begin{equation}
 L_{\mathrm{LP}}=\min_{q\in\mathcal P}\widehat T(q)
 \leq \min_{m\in\mathcal M}\widehat T(m)=J^\star.
 \label{eq:prelim_lp_bound}
\end{equation}
The enlarged feasible set gives a lower bound, while fractional solutions need not represent realizable mappings. PCR uses this bound to screen local mapping domains before integer candidate generation (Section~\ref{sec:lp_mip}).

\subsection{Placement and Timing Closure}

Cell locations and interconnect geometry determine wire resistance and capacitance. Static timing analysis (STA) combines cell models with parasitics estimated from placement or global routing to compute arrival and required times. Setup slack is required time minus arrival time; negative slack indicates a violation. Worst negative slack (WNS) measures the largest setup violation, and total negative slack (TNS) sums negative endpoint slacks.

\subsection{Post-Placement Timing Optimization}

Post-placement timing optimization adjusts cell drive strengths and interconnect loads through sizing, buffering, and related transformations. Lagrangian-relaxation-based sizing~\cite{livramento2013lr,flach2014rsz} optimizes cell sizes under timing constraints; BufFormer learns buffer-tree topology, sizes, and locations~\cite{liang_bufformer_2023}; OpenROAD's Resizer~\cite{ajayi2019openroad} provides cell and interconnect repair operations. PCR complements these techniques by replacing local logic with equivalent mappings, using estimated positions and interconnect delays for search and physical evaluation for acceptance.

\section{Approach}
\label{sec:method}

PCR consists of four stages: (1)~\textbf{timing-driven window extraction}, which selects a violating endpoint and extracts a bounded logic region; (2)~\textbf{mapping formulation and LP screening}, which estimates candidate positions and interconnect delays, models legal mapping choices, and screens for modeled improvement; (3)~\textbf{integer mapping and reconstruction}, which uses MIP to generate concrete candidate netlists; and (4)~\textbf{physical evaluation and iterative update}, which validates candidates through physical checks and global-routing-based timing evaluation before accepting a change. These stages repeat until timing is closed or the search budget is exhausted.

\subsection{Timing-Driven Window Extraction}
\label{sec:vertex}
\label{sec:extraction}
PCR extracts a bounded combinational window rooted at the worst-slack data endpoint, subject to limits on depth, instance count, and boundary inputs. Sequential and macro instances remain outside the window. The extracted context includes interface nets, reference STA timing, external pin loads, and the positions of existing cells, boundary sources, and external sinks. Boundary-source and external-sink positions anchor candidate position estimates. Signals that drive loads outside the window are retained as boundary outputs, preserving their functions and external connections without expanding the window to include all fanout instances.

The extracted logic defines an action graph of legal cut–phase choices, fanin dependencies, and available cell variants. Required outputs induce demands on internal nodes, and compatible actions are selected jointly to form a complete mapping.

\subsection{Mapping Formulation and LP Screening}
\label{sec:uct}
\label{sec:lp_mip}
An action $a\in\mathcal A_i$ specifies a cut and its phase implementation at node $i$; its cell variants specify concrete library gates, including available drive strengths. We jointly select actions and cells because a gate's delay depends on both its own drive and its fanout loads. Let $x_{ia}$ select an action, $d_{ip}$ indicate demand for phase $p$, and $r_i$ indicate whether either phase is demanded. The core selection constraints are
\begin{equation}
 \sum_{a\in\mathcal A_i}x_{ia}=r_i,\quad
 d_{ip}\le r_i\le d_{i0}+d_{i1},\quad
 \sum_a q_{iap}=d_{ip},
 \label{eq:mapping_selection}
\end{equation}
where $q_{iap}=x_{ia}d_{ip}$ indicates that action $a$ serves requested phase $p$. Required output phases have $d_{ip}=1$. Each served action activates its required child phases; conversely, a non-output phase can be demanded only if a selected parent requires it. These constraints exclude unsupported internal gates. Requested phases are translated to physical gate phases according to the action's legal phase implementation. For each such gate, $\sum_c z_{iapc}=g_{iap}$ selects exactly one compatible cell when gate-use indicator $g_{iap}=1$.

\subsubsection{Candidate position and interconnect estimation}
Before each solve, we estimate internal node positions from their connectivity to neighboring nodes and fixed boundary locations. Each position is iteratively moved toward a weighted centroid and projected onto a local region $\Omega$ enclosing the original window and interface positions, expanded by $2\,\mu$m and clipped to the core:
\begin{equation}
 \widetilde{\mathbf p}_i=\Pi_\Omega\!\left(
 \frac{\sum_{j\in\mathcal N_i}w_{ij}\mathbf p_j}
      {\sum_{j\in\mathcal N_i}w_{ij}}\right).
 \label{eq:position_estimate}
\end{equation}
Boundary sources remain fixed. The centroid includes connected neighbors, external sinks, and a weak window-center anchor; external sinks receive greater weight to prevent collapse toward the fanins. An action-specific centroid then combines its cut leaves, downstream neighbors, external sinks, and the estimated root position. These coordinates predict connectivity costs and are not placement decision variables.

For each candidate driver, a minimum spanning tree with Manhattan edge lengths approximates its internal fanout net. For clarity, action and cell indices on the tree quantities are suppressed. Let $\ell_e$ be an edge length, $r_w,c_w$ the per-unit-length resistance and capacitance, and $C_e^\downarrow$ the sink and wire capacitance below the edge, excluding its own wire capacitance. The estimated wire capacitance and worst-sink Elmore delay are
\begin{equation}
 \begin{aligned}
 C_i^w&=c_w\sum_{e\in\mathcal T_i}\ell_e,\\
 D_i^w&=\max_{s\in\mathcal S_i}\sum_{e\in P(i,s)}
 r_w\ell_e\left(C_e^\downarrow+\tfrac12 c_w\ell_e\right).
 \end{aligned}
 \label{eq:wire_estimate}
\end{equation}
The first solve uses possible fanouts and their maximum available input-pin capacitances for this RC estimate. If a sparse mapping is obtained, its selected topology and cell loads refresh the estimates for at most one additional solve. Both solves share the search budget. External output RC and downstream timing remain in the fixed reference context; the tree adds only internal interconnect effects.

\subsubsection{Delay model and relaxation}
We minimize modeled target arrival $\widehat T$. With RC coefficients frozen for the current solve, the delay proxy for a selected driver variant $c$ is
\begin{equation}
 \begin{aligned}
 \widehat d_{i,c}&=d^0_{i,c}+\Delta d_{i,c}^{\mathrm{RC}}
 +R_{i,c}\left(C_i^{\mathrm{ext}}+\sum_j C_j^{\mathrm{in}}z_j\right),\\
 \Delta d_{i,c}^{\mathrm{RC}}&=R_{i,c}C_i^w+D_i^w+D_{i,c}^{\mathrm{in}}.
 \end{aligned}
 \label{eq:method_delay_proxy}
\end{equation}
Here $j$ indexes fanout cell variants; the internal pin load uses their mean input-pin capacitances. For a gate directly driven by window inputs, $D_{i,c}^{\mathrm{in}}$ is the maximum input-wire estimate $r_w L(C_{i,c}^{\mathrm{in}}+c_wL/2)$ over its boundary-source connections of Manhattan length $L$; it is zero otherwise. Reference STA offsets calibrate the local arrival objective, while external pin loads remain fixed. The RC correction is a constant for each variant within a solve, preserving linearity. The model does not predict congestion, obstacle detours, or slew propagation. Conditional arrival constraints propagate delays through selected dependencies. Selecting a driver and its load introduces products of selection variables. Each product $w=uv$ is represented by
\begin{equation}
 0\le w,\qquad w\le u,\qquad w\le v,\qquad w\ge u+v-1.
 \label{eq:mapping_product}
\end{equation}
These inequalities are exact when $u,v$ are binary, so the discrete proxy is a mixed-integer linear model.

We relax the remaining binary domains to $[0,1]$, retaining all selection, dependency, timing, and product constraints. For fractional selections, Eq.~\eqref{eq:mapping_product} bounds their products, yielding the relaxation
\begin{equation}
 L_{\mathrm{LP}}=\min_{\mathcal F_{\mathrm{LP}}}\widehat T
 \ \le\ \min_{\mathcal F_{\mathrm{IP}}}\widehat T,
 \qquad \tau=\widehat T_0-\delta .
 \label{eq:mapping_lp_screen}
\end{equation}
Here $\mathcal F_{\mathrm{IP}}$ is the integer domain under the configured search limits and $\mathcal F_{\mathrm{LP}}$ its relaxation. A certified bound $L_{\mathrm{LP}}>\tau+\epsilon_{\mathrm{cert}}$, or proven infeasibility of the threshold relaxation, rejects the modeled target; $\epsilon_{\mathrm{cert}}$ accounts for numerical tolerance. We use the original starting-mapping score as the screening threshold ($\delta=0$), unchanged across the two solves. The lower bound applies only to the specified candidate domain and the current frozen RC model; it does not bound timing after routing or a subsequent model update.

\subsection{Integer Mapping and Reconstruction}
\label{sec:representatives}
Integer mapping jointly determines the gates and their connections under the current frozen delay model. We define the search domain and representative objectives below; Algorithm~1 specifies their execution order and the optional model refresh. Let $m$ denote a model assignment with action/cell selections $\mathbf b(m)$ and continuous timing variables.

\subsubsection{Restrict the action domain}
Let $\bar x_{ia}$ be the LP value of action $a$ at node $i$. We retain the two highest-ranked actions at each node (or all actions if fewer are available):
\begin{equation}
 \begin{aligned}
 \mathcal A_i'&=\operatorname{Top}_2\{a\in\mathcal A_i;\bar x_{ia}\},\\
 \mathcal F'&=\{m\in\mathcal F_{\mathrm{IP}}:
 x_{ia}=0\ \text{for }a\notin\mathcal A_i'\}.
 \end{aligned}
 \label{eq:restricted_mapping_domain}
\end{equation}
Compatible cell variants remain available. MIP jointly selects actions and cells subject to dependency and phase constraints. Infeasibility and optimality conclusions apply to the restricted domain $\mathcal F'$.

\ifdefined\SingleColumnTableOne\else
\begin{table*}[t]
\centering\caption{Position-aware comparison with PigMap3 on all nine completed pairs. Both methods use detailed placement, setup timing repair, and final detailed placement.}\label{tab:main_comparison}
\fontsize{8}{9.5}\selectfont\setlength{\tabcolsep}{2.5pt}
\begin{tabular*}{0.75\textwidth}{@{\extracolsep{\fill}}lrrrrrrrr@{}}\toprule
 & \multicolumn{4}{c}{PCR (Ours)} & \multicolumn{4}{c}{PigMap3~\cite{pan2026pigmap3}}\\
\cmidrule(lr){2-5}\cmidrule(lr){6-9}
Design & WNS & TNS & Area & Leakage & WNS & TNS & Area & Leakage\\
\midrule\multicolumn{9}{l}{\textit{ICCAD 2024 Problem C~\cite{wu2024contest}}}\\
mempool\_tile\_wrap & \textbf{-100.90} & \textbf{-2994.80} & \textbf{21088.06} & \textbf{2594.1801} & -8722.70 & -32259792.00 & 21224.38 & 2594.2643\\
aes\_256 & \textbf{0.00} & \textbf{0.00} & \textbf{24001.03} & 16.9181 & -61.57 & -356.96 & 24006.67 & \textbf{16.9176}\\
hidden1 & \textbf{-141.79} & \textbf{-7496.61} & \textbf{5627.56} & \textbf{2.8745} & -3049.58 & -348824.22 & 5880.08 & 3.0498\\
hidden5 & \textbf{0.00} & \textbf{0.00} & \textbf{28068.42} & \textbf{26.8309} & -300.52 & -3544.54 & 28095.91 & 26.8359\\
\midrule\multicolumn{9}{l}{\textit{MLCAD 2025~\cite{gopalakrishnan_invited_2025}}}\\
ac97\_top & \textbf{-59.50} & \textbf{-22641.40} & \textbf{1157.00} & 83.6219 & -73.65 & -23640.86 & 1159.43 & \textbf{83.4392}\\
aes\_cipher\_top & \textbf{-22.77} & \textbf{-1669.02} & \textbf{1002.96} & \textbf{30.3782} & -29.31 & -1815.61 & 1003.05 & 30.3805\\
partition\_m$^\dagger$ & -127.15 & \textbf{-4891.43} & \textbf{1920.13} & \textbf{171.4317} & \textbf{-124.29} & -4916.95 & 1920.22 & 171.4467\\
partition\_p & -184.16 & -36540.81 & \textbf{9269.18} & 612.6764 & \textbf{-160.88} & \textbf{-33286.33} & 9303.89 & \textbf{611.4790}\\
partition\_c$^\dagger$ & \textbf{-124.44} & \textbf{-1100.77} & \textbf{23591.66} & \textbf{18102.9051} & -162.33 & -1135.72 & 23593.43 & 18102.9473\\
\midrule
Geo. mean$^*$ & \textbf{33.10} & \textbf{818.82} & \textbf{6972.59} & \textbf{136.4062} & 248.63 & 16927.00 & 7017.36 & 137.2482\\
\midrule
Ratio & \textbf{1.00} & \textbf{1.00} & \textbf{1.000} & \textbf{1.000} & 7.51 & 20.67 & 1.006 & 1.006\\
\bottomrule\end{tabular*}\par\smallskip
\begin{minipage}{\textwidth}\footnotesize WNS/TNS: ps; area: $\mu\mathrm{m}^2$; leakage: $\mu$W. Bold denotes the best displayed value, including ties. Names partition\_m/p/c abbreviate NV\_NVDLA\_partition\_m/p/c. $^\dagger$One remapping update accepted; the other seven cases retain their input mappings. $^*$Timing geometric means use $G=[\prod_{i=1}^{n}(|x_i|+1)]^{1/n}-1$ with $n=9$; area/leakage use ordinary geometric means. Ratio is $G_{\mathrm{method}}/G_{\mathrm{PCR}}$; values above one favor the PCR flow. Lower aggregates are better.\end{minipage}
\end{table*}
\fi

\subsubsection{Establish feasibility and optimize timing}
We first search for an integer assignment $m\in\mathcal F'$ satisfying $\widehat T(m)\le\tau$. Once such an incumbent is found, we solve the timing minimization problem to establish the best modeled arrival in the retained domain:
\begin{equation}
 \widehat T^\star=\min_{m\in\mathcal F'}\widehat T(m),\qquad
 \mathcal S=\{m\in\mathcal F':\widehat T(m)\le\widehat T^\star+\epsilon_T\}.
 \label{eq:integer_timing_cap}
\end{equation}
The feasibility stage can stop upon finding a threshold-feasible incumbent, avoiding a full timing optimization for windows without such an incumbent within the allotted budget. The timing stage establishes a common reference for representative selection. We use $\epsilon_T=0.1$ ps to admit alternatives with nearly equal modeled timing.


\subsubsection{Select two representatives}
Within $\mathcal S$, define $N(m)=\sum_{k\in\mathcal N}z_k(m)$, where $\mathcal N$ indexes non-default cell variants of the retained actions. We first minimize $N(m)$ to obtain a sparse representative $m_1$. We then seek a second representative $m_2$ that differs most from $m_1$ in its action and cell selections:
\begin{equation}
 \begin{aligned}
 m_1&\in\operatorname*{lex\,arg\,min}_{m\in\mathcal S}
       \bigl(N(m),\widehat T(m)\bigr),\\
 m_2&\in\operatorname*{lex\,arg\,min}_{m\in\mathcal S}
       \bigl(-H(m,m_1),\widehat T(m)\bigr),\\
 H(m,m_1)&=\sum_{k:b_k(m_1)=0}b_k(m)
             +\sum_{k:b_k(m_1)=1}\bigl(1-b_k(m)\bigr).
 \end{aligned}
 \label{eq:mapping_representatives}
\end{equation}
The sparse objective minimizes non-default cell selections, not total gate count or changes to the input netlist. The diverse objective maximizes structural difference at similar modeled timing. Both objectives use arrival time as a secondary criterion; remaining sparse ties are resolved deterministically.

\begin{figure}[t]
\hrule\vspace{3pt}
{\raggedright\textbf{Algorithm 1: Candidate generation}\par}
\begin{algorithmic}[1]
\REQUIRE Action graph $\mathcal G$, physical context $\mathcal B$, threshold $\tau$, tolerances $\epsilon_T,\epsilon_{\mathrm{cert}}$
\ENSURE Candidate netlist set $\mathcal C$
\STATE $\mathcal C\gets\varnothing$; $m_{\mathrm{prev}}\gets\bot$
\FOR{$k=0,1$ while search budget remains}
\STATE $\theta_k\gets\operatorname{EstimateRC}(\mathcal G,\mathcal B,m_{\mathrm{prev}})$
\STATE Construct $\mathcal F_{\mathrm{IP}},\mathcal F_{\mathrm{LP}},\widehat T$ with frozen $\theta_k$.
\STATE $\bar m\gets\operatorname{SolveLP}(\min_{\mathcal F_{\mathrm{LP}}}\widehat T)$
\STATE \textbf{if} $\bar m=\bot$ or $\widehat T(\bar m)>\tau+\epsilon_{\mathrm{cert}}$: \textbf{break}
\STATE Construct $\mathcal F'$ using~\eqref{eq:restricted_mapping_domain}.
\STATE $m_0\gets\operatorname{FindFeasible}(\mathcal F'\cap\{\widehat T\le\tau\})$
\STATE \textbf{if} $m_0=\bot$: \textbf{break}
\STATE Solve for $\widehat T^\star$ and construct $\mathcal S$ using~\eqref{eq:integer_timing_cap}.
\STATE \textbf{if} timing optimum is unresolved: \textbf{break}
\STATE Solve for $m_1$ using~\eqref{eq:mapping_representatives}.
\STATE \textbf{if} $m_1=\bot$: \textbf{break}
\STATE $\mathcal C_k\gets\operatorname{ValidateAndDeduplicate}(m_1,\varnothing)$
\STATE Solve for $m_2$ using~\eqref{eq:mapping_representatives}, with $m_1$ fixed.
\IF{$m_2\ne\bot$}
\STATE $\mathcal C_k\gets\operatorname{ValidateAndDeduplicate}(m_2,\mathcal C_k)$
\ENDIF
\STATE \textbf{if} $\mathcal C_k=\varnothing$: \textbf{break}
\STATE $\mathcal C\gets\mathcal C_k$; $m_{\mathrm{prev}}\gets m_1$
\ENDFOR
\RETURN $\mathcal C$
\end{algorithmic}
\vspace{3pt}\hrule
\end{figure}

\subsubsection{Reconstruct the netlists}
$\operatorname{ValidateAndDeduplicate}(m,\mathcal C)$ reconstructs the selected gates and connections, preserves the window interfaces, and retains compatible, realizable mappings with distinct structures or cells. It independently recomputes the original cell/load-model score for consistency checks; search uses the RC-augmented score.

In Algorithm 1, $\bot$ denotes failure to meet a solve's completion criterion: feasibility for $\operatorname{FindFeasible}$ or optimality within solver tolerances for optimization. A timeout does not prove infeasibility. An unresolved diverse representative leaves the sparse candidate available; an incomplete model-refresh round retains the previous completed candidate set.

\ifdefined\SingleColumnTableOne
\begin{table}[t]
\centering\caption{Position-aware comparison with PigMap3~\cite{pan2026pigmap3} on all nine completed pairs. Both methods use detailed placement, setup timing repair, and final detailed placement.}\label{tab:main_comparison}
\fontsize{7}{9}\selectfont\setlength{\tabcolsep}{0.6pt}
\begin{tabular*}{\columnwidth}{@{\extracolsep{\fill}}llrrrr@{}}\toprule
Design & Method & WNS & TNS & Area & Leakage\\
\midrule\multicolumn{6}{l}{\textit{ICCAD 2024 Problem C~\cite{wu2024contest}}}\\
\multirow{2}{*}{\shortstack[l]{mempool\_tile\\\_wrap}} & PCR (Ours) & \textbf{-100.90} & \textbf{-2994.80} & \textbf{21088.06} & \textbf{2594.1801}\\
 & PigMap3 & -8722.70 & -32259792.00 & 21224.38 & 2594.2643\\
\multirow{2}{*}{aes\_256} & PCR (Ours) & \textbf{0.00} & \textbf{0.00} & \textbf{24001.03} & 16.9181\\
 & PigMap3 & -61.57 & -356.96 & 24006.67 & \textbf{16.9176}\\
\multirow{2}{*}{hidden1} & PCR (Ours) & \textbf{-141.79} & \textbf{-7496.61} & \textbf{5627.56} & \textbf{2.8745}\\
 & PigMap3 & -3049.58 & -348824.22 & 5880.08 & 3.0498\\
\multirow{2}{*}{hidden5} & PCR (Ours) & \textbf{0.00} & \textbf{0.00} & \textbf{28068.42} & \textbf{26.8309}\\
 & PigMap3 & -300.52 & -3544.54 & 28095.91 & 26.8359\\
\midrule\multicolumn{6}{l}{\textit{MLCAD 2025~\cite{gopalakrishnan_invited_2025}}}\\
\multirow{2}{*}{ac97\_top} & PCR (Ours) & \textbf{-59.50} & \textbf{-22641.40} & \textbf{1157.00} & 83.6219\\
 & PigMap3 & -73.65 & -23640.86 & 1159.43 & \textbf{83.4392}\\
\multirow{2}{*}{\shortstack[l]{aes\_cipher\\\_top}} & PCR (Ours) & \textbf{-22.77} & \textbf{-1669.02} & \textbf{1002.96} & \textbf{30.3782}\\
 & PigMap3 & -29.31 & -1815.61 & 1003.05 & 30.3805\\
\multirow{2}{*}{partition\_m$^\dagger$} & PCR (Ours) & -127.15 & \textbf{-4891.43} & \textbf{1920.13} & \textbf{171.4317}\\
 & PigMap3 & \textbf{-124.29} & -4916.95 & 1920.22 & 171.4467\\
\multirow{2}{*}{partition\_p} & PCR (Ours) & -184.16 & -36540.81 & \textbf{9269.18} & 612.6764\\
 & PigMap3 & \textbf{-160.88} & \textbf{-33286.33} & 9303.89 & \textbf{611.4790}\\
\multirow{2}{*}{partition\_c$^\dagger$} & PCR (Ours) & \textbf{-124.44} & \textbf{-1100.77} & \textbf{23591.66} & \textbf{18102.9051}\\
 & PigMap3 & -162.33 & -1135.72 & 23593.43 & 18102.9473\\
\midrule
\multirow{2}{*}{Geo. mean$^*$} & PCR (Ours) & \textbf{33.10} & \textbf{818.82} & \textbf{6972.59} & \textbf{136.4062}\\
 & PigMap3 & 248.63 & 16927.00 & 7017.36 & 137.2482\\
\midrule
\multirow{2}{*}{Ratio} & PCR (Ours) & \textbf{1.00} & \textbf{1.00} & \textbf{1.000} & \textbf{1.000}\\
 & PigMap3 & 7.51 & 20.67 & 1.006 & 1.006\\
\bottomrule\end{tabular*}\par\smallskip
\begin{minipage}{\columnwidth}\footnotesize WNS/TNS: ps; area: $\mu\mathrm{m}^2$; leakage: $\mu$W. Bold denotes the best displayed value, including ties. Names partition\_m/p/c abbreviate NV\_NVDLA\_partition\_m/p/c. $^\dagger$One remapping update accepted; the other seven cases retain their input mappings. $^*$Timing geometric means use $G=[\prod_{i=1}^{n}(|x_i|+1)]^{1/n}-1$ with $n=9$; area/leakage use ordinary geometric means. Ratio is $G_{\mathrm{method}}/G_{\mathrm{PCR}}$; values above one favor the PCR flow. Lower aggregates are better.\end{minipage}
\end{table}
\fi

\subsection{Physical Evaluation and Iterative Update}
\label{sec:eval}
 Predicted coordinates serve only the delay model. Each reconstructed candidate is applied to a separate copy of the current placed design, where actual placement and legalization determine its physical locations. Detailed placement legalizes the modified region, followed by placement checking, global routing, parasitic estimation, and STA. The candidate and reference are evaluated using the same timing constraints and physical settings.

 A candidate is accepted when WNS and TNS do not degrade beyond tolerance, at least one improves, and the composite cost decreases. The tolerances and cost function are specified in Section~\ref{sec:whole_design_protocol}. If the candidate passes, we retain its design, recompute the search context, and restart endpoint selection from the worst slack. Otherwise, we retain the previous design, try any remaining representative, and then advance to the next endpoint. Iteration stops when setup timing is closed, the window limit is reached, or the wall-time budget expires.

\section{Experimental Results and Discussion}
\label{sec:results}

\subsection{Experimental Setup}
Experiments run on a Linux cluster with Intel Xeon processors, using single-threaded workers. Both benchmark suites use ASAP7 libraries~\cite{ASAP7}. OpenROAD~\cite{ajayi2019openroad} provides the physical-design backend for placement, timing repair, global routing, parasitic estimation, and static timing analysis. We compare PCR with OpenROAD's genetic-algorithm and simulated-annealing based restructuring methods (\texttt{resynth\_genetic} and \texttt{resynth\_annealing}), and with PigMap3~\cite{pan2026pigmap3}. All methods use the same initial checkpoint, libraries, timing constraints, and final evaluation settings for each design.

We evaluate benchmarks from the 2024 ICCAD Contest Problem C~\cite{wu2024contest} and the 2025 MLCAD Contest~\cite{gopalakrishnan_invited_2025}. Table~\ref{tab:main_comparison} lists all nine completed position-aware pairs with PigMap3~\cite{pan2026pigmap3}; Table~\ref{tab:uniform_comparison} lists all seven completed uniform-flow comparisons with annealing and genetic. Each table uses a common case set for its methods and reports WNS, TNS, standard-cell area, and leakage power.

PCR uses extraction limits of depth six, 60 instances, and 40 boundary inputs, with an endpoint pool of 256. Each run attempts at most 20 windows within $\max(1800,\lceil2t_{\mathrm{ref}}+600\rceil)$ seconds, where $t_{\mathrm{ref}}$ is reference evaluation time. The LP/MIP selection rules and timing tolerances are specified in Section~\ref{sec:method}. Resynthesis and physical-stage timeouts are 7200 and 14400 s, respectively; search budgets are method-specific.

\label{sec:whole_design_protocol}
During local restructuring, routed candidate acceptance allows at most $0.001$ ps degradation in WNS/TNS, requires improvement exceeding $0.001$ ps in at least one, and requires a decrease exceeding $10^{-6}$ in
\begin{equation}
C=P_{\mathrm{leak}}+10|\min(\mathrm{TNS},0)|+20V_{\mathrm{slew}}+20V_{\mathrm{cap}},
\label{eq:whole_design_cost}
\end{equation}
with leakage, TNS, slew violation, and capacitance violation expressed in $\mu$W, ns, ns, and fF. 

\begin{table*}[t]
\centering\caption{Uniform-flow comparison on all seven cases completed by all three methods. Every method uses global and detailed placement before and after setup timing repair.}\label{tab:uniform_comparison}
\fontsize{8}{9.5}\selectfont\setlength{\tabcolsep}{2.5pt}
\begin{tabular*}{0.9\textwidth}{@{\extracolsep{\fill}}lrrrrrrrrrrrr@{}}\toprule
 & \multicolumn{4}{c}{PCR (Ours)} & \multicolumn{4}{c}{Annealing} & \multicolumn{4}{c}{Genetic}\\
\cmidrule(lr){2-5}\cmidrule(lr){6-9}\cmidrule(lr){10-13}
Design & WNS & TNS & Area & Leakage & WNS & TNS & Area & Leakage & WNS & TNS & Area & Leakage\\
\midrule\multicolumn{13}{l}{\textit{ICCAD 2024 Problem C~\cite{wu2024contest}}}\\
partition\_m & \textbf{-236.59} & \textbf{-10784.62} & 2759.48 & 2.7394 & -438.20 & -21553.12 & \textbf{2073.74} & \textbf{1.9454} & -485.89 & -23138.89 & 2302.42 & 2.1421\\
partition\_p & \textbf{-172.12} & \textbf{-21405.49} & 10850.60 & 6.6562 & -391.17 & -100520.52 & \textbf{10587.47} & \textbf{6.4325} & -375.68 & -98190.02 & 10675.18 & 6.5484\\
aes\_256 & \textbf{-427.70} & \textbf{-9144.16} & \textbf{24001.04} & \textbf{16.9181} & \textbf{-427.70} & \textbf{-9144.16} & \textbf{24001.04} & \textbf{16.9181} & \textbf{-427.70} & \textbf{-9144.16} & \textbf{24001.04} & \textbf{16.9181}\\
hidden1 & \textbf{-73.06} & -2843.29 & \textbf{5627.71} & \textbf{2.8747} & -112.18 & -4764.22 & 5682.15 & 2.9453 & -125.23 & \textbf{-1298.82} & 5706.71 & 2.9664\\
hidden5 & \textbf{-156.44} & -4817.11 & \textbf{28104.82} & \textbf{26.8765} & -157.45 & \textbf{-4138.66} & \textbf{28104.82} & \textbf{26.8765} & -157.45 & \textbf{-4138.66} & \textbf{28104.82} & \textbf{26.8765}\\
\midrule\multicolumn{13}{l}{\textit{MLCAD 2025~\cite{gopalakrishnan_invited_2025}}}\\
ac97\_top & \textbf{-65.50} & \textbf{-24721.82} & \textbf{1156.16} & 83.4442 & -162.46 & -142236.77 & 1271.77 & \textbf{66.7008} & -159.43 & -123481.26 & 1291.86 & 72.3576\\
aes\_cipher\_top & \textbf{-41.80} & \textbf{-2806.58} & \textbf{1009.81} & \textbf{30.4120} & -301.08 & -39003.28 & 1292.66 & 35.4963 & -236.81 & -30209.31 & 1266.97 & 31.8558\\
\midrule
Geo. mean$^*$ & \textbf{127.71} & \textbf{7949.77} & \textbf{5393.12} & 12.9319 & 251.47 & 21563.35 & 5425.30 & \textbf{12.1761} & 248.39 & 17038.68 & 5513.40 & 12.3416\\
\midrule
Ratio & \textbf{1.00} & \textbf{1.00} & \textbf{1.000} & 1.000 & 1.97 & 2.71 & 1.006 & \textbf{0.942} & 1.94 & 2.14 & 1.022 & 0.954\\
\bottomrule\end{tabular*}\par\smallskip
\begin{minipage}{\textwidth}\footnotesize WNS/TNS: ps; area: $\mu\mathrm{m}^2$; leakage: $\mu$W. Bold denotes the best displayed value, including ties. Names partition\_m/p/c abbreviate NV\_NVDLA\_partition\_m/p/c. All seven cases retain their input mappings during remapping. $^*$Timing geometric means use $G=[\prod_{i=1}^{n}(|x_i|+1)]^{1/n}-1$ with $n=7$; area/leakage use ordinary geometric means. Ratio is $G_{\mathrm{method}}/G_{\mathrm{PCR}}$; values above one favor the PCR flow. Lower aggregates are better.\end{minipage}
\end{table*}

\subsection{Experimental Flow}
We organize comparisons by matched physical postprocessing. PCR and PigMap3~\cite{pan2026pigmap3} are compared in the \emph{position-aware flow}: detailed placement, setup timing repair, and final detailed placement. Annealing and genetic are compared with PCR in the \emph{uniform flow}: global and detailed placement both before and after timing repair. These settings assess placement-preserving optimization and optimization followed by common global re-placement, respectively. For each method, the two flows reuse the same mapping output. Repair uses \texttt{repair\_timing -setup}, allowing sizing, buffering, pin swapping, and cloning. Final checkpoints undergo legalization and global-routing-based timing evaluation.

\subsection{Results Analysis}
Each table reports geometric means and ratios over its own completed case set. Timing aggregates use add-one smoothing to include zero violations. The two tables have different case sets and physical flows, so their aggregate values are not directly comparable.

\subsubsection{Position-Aware Comparison with PigMap3}
The PCR flow achieves better WNS on seven of nine cases and better TNS on eight. It closes timing on \texttt{aes\_256} and \texttt{hidden5}, and gives lower area on all nine cases. The large aggregate timing advantage reflects both these zero-violation results and large PigMap3 violations on \texttt{mempool\_tile\_wrap} and \texttt{hidden1}; its magnitude depends on the smoothing convention.

Only MLCAD \texttt{partition\_m} and \texttt{partition\_c}, marked in Table~\ref{tab:main_comparison}, accept a remapping update. Before timing repair, these updates improve WNS by 48.61 and 1976.43 ps relative to their respective initial designs. After the complete flow, \texttt{partition\_c} improves WNS/TNS over PigMap3 by 37.90/34.95 ps, with slightly smaller area and leakage. On \texttt{partition\_m}, TNS violation decreases by 0.52\%, while WNS worsens by 2.86 ps. The remaining seven cases retain their input mappings, so the table's overall advantage measures complete-flow outcomes rather than isolated remapping gains.

\subsubsection{Uniform Comparison with Annealing and Genetic}
Across seven completed cases, the PCR flow reduces geometric-mean WNS and TNS violation magnitudes by 48.6--49.2\% and 53.3--63.1\%, respectively, relative to annealing and genetic. It has the best WNS on all seven cases, including a tie on \texttt{aes\_256}, and the best TNS on five, also including that tie. Area and leakage exhibit design-dependent tradeoffs, as shown in Table~\ref{tab:uniform_comparison}.



\section{Conclusion}
\label{sec:conclusion}

We presented PCR, a physically contextualized local restructuring method for post-placement timing optimization. PCR extracts bounded windows around timing-critical endpoints while preserving their interfaces to the surrounding design. Within each window, PCR predicts gate positions and interconnect RC to jointly optimize cut, phase, and cell choices under a delay model frozen for each solve. The selected topology can refresh this model once; an LP relaxation screens for modeled improvement and guides action-domain restriction, followed by integer optimization to select sparse and structurally diverse candidate mappings. Candidates are reconstructed, legalized, and evaluated using global-routing parasitics before acceptance. Accepted updates guide subsequent window selection, enabling iterative optimization within a fixed computation budget. Experimental results show that the complete PCR flow outperforms PigMap3~\cite{pan2026pigmap3} with 86.7\% and 95.2\% reductions in geometric-mean WNS and TNS violation magnitudes under the position-aware flow, and genetic and annealing with reductions of 48.6--49.2\% and 53.3--63.1\% under the uniform flow. It outperforms PigMap3 in WNS on seven of nine cases and each resynthesis baseline on six of seven cases, with one tie.

\vspace{-2mm}



\clearpage
\balance
\bibliographystyle{IEEEtran}
\bibliography{ref/refs}

\end{document}